# HST IMAGES OF NEARBY LUMINOUS QUASARS II: RESULTS FOR EIGHT QUASARS AND TESTS OF THE DETECTION SENSITIVITY[1]

John N. Bahcall, Sofia Kirhakos

Institute for Advanced Study, School of Natural Sciences, Princeton, NJ 08540

and

Donald P. Schneider

Department of Astronomy and Astrophysics, The Pennsylvania State University, University Park, PA 16802

and

Institute for Advanced Study, School of Natural Sciences, Princeton, NJ 08540

astro-ph/9501018    6 Jan 95

---

# ABSTRACT

HST observations of eight intrinsically luminous quasars with redshifts between 0.16 and 0.29 are presented. Seven companion galaxies brighter than $M_V = -16.5$ ($H_0 = 100$ kms$^{-1}$Mpc$^{-1}$, $\Omega_0 = 1.0$) lie within a projected distance of 25 kpc of the quasars; three of the companions are located closer than 3″ (6 kpc projected distance) from the quasars, well within the volume that would be enclosed by a typical $L^*$ host galaxy. The observed association of quasars and companion galaxies is statistically significant and may be an important element in the luminous-quasar phenomenon.

Evidence for candidate host galaxies is presented for the three most promising cases: PG 1116+215, 3C 273, and PG 1444+407, but additional observations are required before the characteristics of the candidate hosts can be regarded as established.

Upper limits are placed on the visual-band brightnesses of representative galactic hosts for all of the quasars. These limits are established by placing galaxy images obtained with HST underneath the quasars and measuring at what faintness level the known galaxies are detected. On average, the HST spirals would have been detected if they were one magnitude fainter than $L^*$ or brighter and the early-type galaxies could have been detected down to a brightness level of about $L^*$, where $L^*$ is the Schechter characteristic luminosity of field galaxies. Smooth, featureless galaxy models (exponential disks or de Vaucouleurs profiles) are fit to the residual light after a best-fitting point source is subtracted from the quasar images. The results show that smooth host galaxies brighter than, on average, about $L^*$, would have been detected. These upper limits, or possible detections, are consistent with, for example, the eight luminous quasars studied in this paper occurring in host galaxies that have a Schechter luminosity function with a lower-cutoff of in the range $0.01L^*$ to $0.1L^*$.

Tests are performed to determine if our failure to detect luminous host galaxies could be an artifact caused by our analysis procedures. These tests include comparing the measured PSF for our HST observations with the PSFs used in previous ground-based studies of host galaxies, measuring the fluctuations in the sky signals that were subtracted from the quasar images, evaluating empirically the effects of using different stellar point spread functions in the analysis, carrying out the subtraction of the stellar (nuclear) source in different ways, creating and analyzing artificial AGNs with known surface brightnesses, and fitting the observed quasar light to an analytic model that includes a host galaxy. Our analysis procedures successfully pass all of these tests.

*Subject headings:* quasars: individual (PG 0953+414, PG 1116+215, PG 1202+281, PKS 1302−102, 3C 273, PG 1307+085, PG 1444+407, 3C 323.1)

## 1. INTRODUCTION

It has long been recognized that quasars might reside in galaxies and that the quasar phenomenon might be strongly influenced by the presence of companion galaxies. Depending upon your theoretical bias, you may expect to find quasars in old galaxies, in young galaxies, or in interacting galaxies, all possibilities that have been advocated in the published literature. Very close companions might provide fuel for, or initiate processes that lead to, the quasar phenomenon.

Over the past two decades there have been a large number of ground-based observational programs in which the properties of host galaxies of quasars were investigated. In fact, a consensus view has been developed that quasars reside in luminous galaxies, with the intrinsically brightest quasars residing in the most luminous galaxies. Moreover, many authors have concluded that radio bright quasars reside in elliptical galaxies and radio faint quasars reside in spiral galaxies. Some representative papers providing evidence for these views are Kristian (1973), Wyckoff, Wehinger, & Gehren (1981), Hutchings et al. (1982), Gehren, et al. (1984), Heckman et al. (1984), Malkan (1984), Malkan, Margon, & Chanan (1984), Boroson, Persson, & Oke (1985), Smith et al. (1986), Hutchings (1987), Stockton & MacKenty (1987), Yee (1987), Hutchings, Janson, & Neff (1989), Romanishin & Hintzen (1989), Véron-Cetty & Woltjer (1990), Hutchings & Neff (1992), Dunlop et al. (1993), and McLeod & Rieke (1994a,b). The analyses presented in these pioneering studies are difficult because of atmospheric seeing; the light from the bright central (nuclear) sources may be a few magnitudes brighter than the total emission from the host galaxies.

Some of the papers that have previously discussed the companions of low-redshift quasars include Bahcall, Schmidt, and Gunn (1969), Bahcall and Bahcall (1970), Gunn (1971), Robinson and Wampler (1972), Burbidge and O'Dell (1973), Stockton (1978), Hutchings et al. (1982), Heckman et al. (1984), Malkan, Margon, & Chanan (1984), Green & Yee (1984), and Yee (1987). The most dramatic results presented in the present paper refer to very close galaxy companions that are not visible in previously-obtained ground-based images.

The excellent optical characteristics of the repaired HST make possible improved observational studies of the hosts of small-redshift quasars. The characterization of faint galactic material (the host galaxy) in the vicinity of a bright stellar source (the quasar) can be done with increased reliability using the HST to minimize the possibility that quasar light is mistaken for diffuse galactic light. Close galactic companions (within 2″ or 3″ of the quasar nucleus) may also be most convincingly detected with HST.

We describe in this paper the systematic results that have been obtained for the first eight objects in a sample of 18 of the intrinsically most luminous ($M_V < -22.9$, for $H_0 = 100$ kms$^{-1}$Mpc$^{-1}$, $\Omega_0 = 1.0$) nearby ($z < 0.30$) radio-quiet and radio-loud quasars selected from the Véron-Cetty & Véron (1993) catalog. Images of the first four of the quasars (PG 0953+414, PG 1116+215, PG 1202+281 (GQ Com), and PG 1307+085) were presented in Bahcall, Kirhakos, & Schneider (1994, Paper I); images of four additional quasars (3C 273, PKS 1302−102, PG 1444+407, 3C 323.1) have not previously been discussed (hereafter, " the *new* quasars "). Hutchings et al. (1994) reported on observations of two AGNs imaged with the planetary camera of WFPC2. In the absence of an empirical stellar PSF, they used a model PSF and image restoration to analyze their images and to compare with ground-based data.

Our principal results for all eight quasars and their hosts are summarized in Table 1, which lists the following quantities for each object: the date observed, the redshift and the apparent $V$-magnitude (from Véron-Cetty & Véron 1993), the distance in kpc that corresponds to an angular separation of one arcsec as seen from earth, the absolute $V$-magnitude, and the range of galaxy magnitudes at which eight representative galaxies (selected from HST WFC2 observations with the same filter) could be detected when artificially placed underneath the quasar images. The last column gives the average limiting absolute magnitude at which the eight representative galaxies could have been detected as hosts. Accurate (1″) coordinates and multicolor CCD photometry for six of the eight quasars are given in Kirhakos et al. (1994). Three of the quasars, 3C 273, PKS 1302−102, and 3C 323.1, are radio loud; the other five quasars are radio quiet (Kellermann et al. 1989).

The eight quasars studied in this paper have an average absolute visual magnitude of $<M_V> = -23.9$, approximately two magnitudes brighter than the brightest galaxies of rich clusters (Hoessel & Schneider 1985; Postman & Lauer 1995) and about 3.4 magnitudes brighter than the characteristic (Schechter-) magnitude for field galaxies (Schechter 1976; Kirshner et al. 1983; Efstathiou, Ellis, & Peterson 1988). The average apparent brightness for the quasars is $<V> = 15.2$. At the typical redshift of the quasars, ($z \sim 0.2$), an $L^*$ galaxy would have an apparent magnitude of $V \approx 18.6$.

Table 1: Sample of Completed Cycle 4 Quasars and Detectable Hosts

| Object | Date 1994 | $z$ | kpc per[a] arcsec | $V$ | $M_V(\text{QSO})$[a] | $m_{\text{lim}}(\text{host})$ F606W | $\langle M_{\text{lim}}(\text{host})\rangle$[a] F606W |
|---|---|---|---|---|---|---|---|
| PG 0953+414 | 3 Feb | 0.239 | 2.4 | 15.3 | −24.1 | ≳ 18.9 − 20.3 | −19.9 |
| PG 1116+215 | 8 Feb | 0.177 | 1.9 | 15.0 | −23.7 | ≳ 18.3 − 19.5 | −19.3[b] |
| PG 1202+281 | 8 Feb | 0.165 | 1.8 | 15.6 | −23.0 | ≳ 18.4 − 20.2 | −19.0 |
| 3C 273 | 5 Jun | 0.158 | 1.8 | 12.8 | −25.7 | ≳ 16.7 − 18.3 | −21.1[b] |
| PKS 1302−102 | 9 Jun | 0.286 | 2.7 | 15.2 | −24.6 | ≳ 18.3 − 20.1 | −20.1 |
| PG 1307+085 | 4 Apr | 0.155 | 1.8 | 15.3 | −23.1 | ≳ 18.4 − 20.1 | −19.4 |
| PG 1444+407 | 27 Jun | 0.267 | 2.6 | 15.7 | −24.0 | ≳ 18.6 − 20.3 | −20.5[b] |
| 3C 323.1 | 10 Jun | 0.266 | 2.6 | 16.7 | −22.9 | ≳ 18.7 − 20.2 | −19.0 |

[a] Computed for $\Omega_0 = 1.0$ and $H_0 = 100$ km s$^{-1}$Mpc$^{-1}$. In this cosmology, brightest cluster galaxies have $M_V \approx -22.0$ and the characteristic (Schechter-) magnitude for field galaxies is $M_V^* = -20.5$.
[b] Candidate host galaxy (see § 5).

This paper is organized as follows. In § 2, we describe the observational procedures, including the determination of the stellar point spread function (PSF). We outline, in § 3, the method by which we set limits on the brightnesses of representative HST-observed host galaxies. This section describes how we subtract the stellar PSF (§ 3.1), displays a catalog of observed HST galaxies that are used as simulated host galaxies (§ 3.2), and discusses the determination of the faintest magnitudes at which we could detect host galaxies underneath each observed quasar (§ 3.3). We discuss in § 4 fits of the residual diffuse light to smooth, featureless galaxy models, either an exponential disk or a de Vaucouleurs profile. We present in § 5 the characteristics of the candidate host galaxies that we have found for PG 1116+215, 3C 273, and PG 1444+407. The HST images reveal separate galaxies whose projected separations from the centers-of-light of the quasar images are very small; these companion galaxies are discussed in § 6 and their properties are summarized in Table 6. McLeod & Rieke (1994b) have reported $H-$band detections of host galaxies for all eight of the quasars we discuss in this paper and have given sufficient detail that we can simulate the host galaxies they describe. We compare, in § 7, our limits on the $V$–band brightnesses of simulated McLeod-Rieke galaxies with the reported $H$–band galaxy magnitudes and infer a minimum $V - H$ color for the host galaxies if both observational analyses, theirs and ours, are correct. In § 8, we attempt to answer the question: What have we done wrong? Why have we failed to detect the luminous host galaxies reported by previous investigators who used data obtained with ground-based telescopes? We discuss in § 8 the difference between the PSF measured for our HST images and the PSFs used in previous ground-based studies of host galaxies (§ 8.1), the sky subtraction with the HST images (§ 8.2), the results of experiments in which the images of different stars (observed at different positions in the camera or through different filters) were subtracted from each other(§ 8.3), a comparison of separate analyses performed when the subtraction of a stellar point source was optimized on the diffraction spikes alone or on an azimuthally average light intensity, (§ 8.4), and a successful analysis of simulated AGNs (§ 8.5). We summarize and discuss our principal conclusions in § 9.

## 2. OBSERVATIONS

Most of the details of the observational procedures are given in Paper I; we summarize here the main features of the observations. We describe in § 2.1 the main features of the quasar images and present in § 2.2 the empirically-determined stellar point spread function.

### 2.1. Quasar Images with F606W

The quasars were observed with the Wide Field Camera-2 (WFC2) through the F606W filter (see Burrows 1994a for a description of the instrument and the response curve of the filter). The F606W filter is similar to the $V$ bandpass but is slightly redder; the mean wavelength and FWHM of the F606W system response are 5940 Å and 1500 Å, respectively. This filter was chosen because of its high throughput, which was the primary consideration for the present exploratory program. At a given angular radius, the scattered light in the WFC2 is about five times less than the scattered light in the planetary camera WFPC2 (see Krist & Burrows 1994). We chose to use the WFC instead of the PC in the original formulation of this program because of the likelihood that the host galaxies might have low surface brightnesses that extended over areas large compared to the WFC resolution (0.1″ or about 0.2 kpc).

The STScI photometric calibration was verified by comparing, in quasar fields obtained at Palomar by Kirhakos et al. (1994), HST galaxy magnitudes with ground-based photometry of the same galaxies. For the purposes of this paper, the $V$ and the $F606W$ photometric bands are sufficiently similar (Bahcall et al. 1994; Holtzman 1994) that we will sometimes use $V$ and $m_{F606W}$ interchangeably.

All quasar images were located at a distance of $5'' \pm 1.5''$ from the center of Wide Field CCD 3 and, except for PG 1307+085, were observed for three separate exposures of 1100 s, 600 s, and 100 s; the exposures for PG 1307+085 lasted 1400 s, 500 s, and 200 s. The image scale of the Wide Field detectors is $0.0966''$ pixel$^{-1}$ (Trauger et al. 1994), which means that the data are severely undersampled. The innermost regions of the quasar images are saturated in all of the exposures out to a radius $\approx 0.3''$. Typically, there were about 9 saturated pixels in the shortest exposures and 50 saturated pixels in the longest exposures. The exposures for 3C 273 contained about 200 saturated pixels in the longest exposure and about 30 saturated pixels in the shortest exposure. The shortest exposure was saturated in the inner three pixels.

The initial data processing (bias frame removal and flat-field calibration) was performed at the Space Telescope Science Institute with their standard software package. The individual images of each quasar were aligned to better than 0.3 pixel; this made it easy to identify and eliminate cosmic ray events. The flat-field corrections were based upon pre-flight calibrations; these calibrations remove the small-scale (few pixel) sensitivity variations. However, there are apparent large-scale sensitivity variations present in the calibrated data. The typical signal and rms of the noise of the sky in the long exposures (in detected photons) are 90 and 14, respectively. The sky level corresponds to a surface brightness of approximately 22.8 mag per sq arc sec. The observed noise is slightly larger than expected from a combination of shot noise and CCD readout noise (7 electrons per pixel). The formal detection limits for extended objects, calculated from these numbers, are extremely faint. However, one limiting factor for the detection of galaxies is the imperfect match between the data and the pre-flight flat-field calibrations.

Figure 1 displays the images of the eight quasars, with the only additional processing beyond that performed by STScI being the removal of cosmic rays. No luminous host galaxies are visible in the raw data, which contrasts with what was expected based upon our studies of simulated AGNs with bright host galaxies (see § 8.5).

## 2.2. The Stellar Point Spread Function

The point spread function (PSF) was determined using a set of images of an isolated $V = 10.5$ mag star (F141) in M67 that was placed on the corners of a box $3''$ on a side. The box was centered at a point displaced $2.5''$ away from the central pixel of Wide Field CCD 3 in each of the two orthogonal directions. At each corner of the box, a 12 s exposure was taken. In addition, a 2.6 s and a 70 s exposure were obtained at one corner of the box. The M67 observations, which were obtained on 5 June 1994 (the same day as the observations of 3C 273 were taken), cover the saturation range present in the quasar images. In these calibration images, the shortest exposures were saturated in the inner $0.2''$, the intermediate exposures were typically saturated to about $0.3''$, and the longest exposures were saturated out to $0.6''$.

Figure 2 and Figure 3 show the point spread function of the HST produced by the F606W filter; this PSF was constructed by combining the short, intermediate, and long exposures of M67-F141. Figure 2 displays the two-dimensional empirical PSF generated from the three exposures of different lengths of the star F141 in M67. The two dimensional profile is slightly asymmetric; the radial spokes are probably due to scattering within the instrument (see Krist & Burrows 1994). Figure 3 shows the azimuthal average of the empirical PSF for F141 and the F606W PSF calculated by Burrows (1994b).

The extended low surface brightness that is apparent in Figure 2 and Figure 3 constitutes the principal difficulty in investigating faint host galaxies of quasars. Even with the excellent PSF of HST, if one fails to subtract precisely the stellar PSF from a quasar image, the residual stellar light from the imperfect subtraction may simulate the appearance of a host galaxy. The actual PSF at a fixed position in the camera is somewhat time-dependent since the focus changes due to desorption and to breathing modes of the telescope system. We will discuss in § 8.3 the limits on the detectability of faint hosts that result from imperfect PSF subtraction.

The PSF determined using F141 is not optimal for removing the nuclear light from quasars, since F141 is considerably redder than the low redshift quasars. Gilliland et al. (1991) measured the following colors for F141: $V = 10.49$, $B-V = +1.11$, $U-B = +1.02$, and $V-R = +0.58$ (see also Eggen & Sandage 1964; Chevalier & Ilovaisky 1991). The Véron-Cetty & Véron (1993) catalog gives accurate colors for five of the eight quasars discussed in this paper. The average quasar colors are: $B - V = +0.05$ and $U - B = -0.86$. Since the measured PSF depends upon color, the empirical PSF constructed using images of F141 is not optimal for subtracting nuclear light from quasars. Empirically, however, the color dependence of the stellar PSF does not cause large errors in the subtraction process (see § 8.3).

## 3. BRIGHTNESS LIMITS ON ORDINARY HOST GALAXIES

We established the brightness upper limits, given in Table 1, on host galaxies of the eight quasars in our sample by artificially placing a series of observed and simulated galaxies underneath the quasar images. We then determined, with both objective and subjective (visual inspection of the data) techniques, the surface brightness levels at which the galaxies are detectable. The detailed procedure used to establish a limiting brightness for each quasar-galaxy pair is described in this section.

The first stage of the analysis, described in § 3.1, consists of subtracting a stellar PSF from the observed quasar image. We then construct a catalog, presented in § 3.2, of observed HST images that span the morphological types of host galaxies. The concluding step, described in § 3.3, is the determination of the limiting magnitude at which each of the putative host galaxies in our catalog of HST images could be detected underneath each of the eight observed quasars.

### 3.1. Subtracting the Stellar PSF

The best fit of the stellar PSF to the quasar image was determined with the downhill simplex method (described in Press et al. 1986) using the $\chi^2$ calculated in relatively narrow regions centered on the diffraction spikes. There are in principle three input parameters to the fit: the central location of the quasar (two coordinates) and the peak brightness of the PSF. However, the location of the quasar could be determined accurately by eye (typically to 0.5 pixel, i.e., $0.05''$) by adjusting the positions of the residuals from the diffraction spikes. In practice, we performed subtractions in which the central position was located by eye and, independently, was located by the simplex method; the two procedures gave indistinguishable results for the residual diffuse emission. Pixels that were saturated in either the quasar data or the PSF data, or that showed a large disagreement between the PSF and the quasar data ($3\sigma$ from the mean deviation), were eliminated from the $\chi^2$ fitting.

Figure 4 shows the result of subtracting the composite best-fit stellar image, Figure 2, (of M67 calibration star F141) from the long-exposure images of each of the eight quasars. Each panel in Figure 4 is $20'' \times 20''$. The contrast has been set to emphasize low-surface brightness features; in the long exposures, the quasar images are saturated to a radius of $\approx 0.6''$.

For the images shown in Figure 4, the intensity of the subtracted stellar images was adjusted visually so as to best eliminate the diffraction spikes. In an independent procedure, azimuthal averages of the residual intensity differences between the quasar and the stellar PSF were minimized using a $\chi^2$-routine. The two methods gave quantitatively similar results (see § 8.4).

### 3.2. Images of test galaxies

In order to quantify our detection limits for faint nebulosity, we established a catalog of representative images of individual galaxies that were added to the HST quasar images. Except for special tests, we used images of galaxies that were observed with the Wide Field CCDs of the WFPC2 through the F606W filter. We also used simulated test galaxies (both exponential disks and de Vaucouleurs spheroids). Unless explicitly specified otherwise, all of the results in this paper refer to the catalog of HST-observed galaxy images.

Figure 5 shows the collection of eight observed galaxy images that were placed underneath the quasars and whose whose surface brightnesses were scaled in order to determine empirically our sensitivity to extended nebulosity centered on the quasar. The first six images appear in Figure 2 of Paper I; they are reproduced here for the reader's convenience.

All eight of the galaxies in Figure 5 were taken from HST observations in this program; they represent a variety of galaxy types. Figure 5a represents a pair of interacting spiral galaxies, Figure 5b is an edge-on spiral, Figure 5c is a ringed spiral seen face on, Figure 5d is a smooth elliptical, Figure 5e is an S0 or an Sa, and Figure 5f appears to be a disturbed barred spiral. The last two panels, Figure 5g and Figure 5h, contain pairs of interacting galaxies.

The exponential scale lengths of the three spiral galaxies, Figure 5a,b,f are, respectively, $0.6''$, $0.6''$, and $0.8''$. We also used in the tests described below images expanded by a linear factor of three for the two spiral galaxies shown in Figure 5b,f. Therefore, the range of exponential scale lengths for the galaxies considered here are, at the redshifts of the quasars, between 1 kpc and 5 kpc. The exponential scale length of the Galaxy is about 3.5 kpc (Bahcall 1986), a value which also describes well the average exponential scale length of other local galaxies and disk-dominated galaxies out to redshifts of order 0.8 (see Mutz et al. 1994)

The de Vaucouleurs effective radii for the three elliptical galaxies shown in Figure 5c,d,e are, respectively, 3.2″, 4.4″, and 1.2″. At the redshifts of the quasars considered in this paper, the effective radii vary between 2 kpc and 9 kpc. For the Galaxy, the effective radius of the spheroid is about 2.7 kpc (Bahcall 1986). It is difficult to define uniquely a length scale for the interacting galaxies shown in Figure 5g,h; the major axes of these systems are, respectively, 9″ and 6″, corresponding to distances of order 12 to 18 kpc.

Table 2 summarizes the characteristic length scales for the eight test galaxies shown in Figure 5. The exponential scale length are given in arcseconds and, for $z = 0.2$, in kpc.

Table 2: Characteristic length scales of the test galaxies

| Galaxy | Scale Length ″ (kpc, $z = 0.2$) | | Galaxy | Scale Length ″ (kpc, $z = 0.2$) | |
|---|---|---|---|---|---|
| 2a | 0.6[a] | (1.3)[a] | 2e | 1.2[b] | (2.5)[b] |
| 2b | 0.6[a] | (1.2)[a] | 2f | 0.8[a] | (1.7)[a] |
| 2c | 3.2[b] | (6.7)[b] | 2g | 1.7[a] | (3.6)[a] |
| 2d | 4.4[b] | (9.2)[b] | 2h | 2.3[a] | (4.9)[a] |

[a] Exponential scale length

[b] de Vaucouleurs effective radius

We have measured from other HST images the angular sizes of galaxies in a cluster at redshift 0.17 and verified that the cluster galaxies have similar angular sizes to the galaxies that were chosen to appear in Figure 5 and which were used in our sensitivity tests (see below).

### 3.3. The Faintest Detectable Test Galaxies

To determine the limiting apparent magnitude at which a specific test galaxy could be detected, the specified galaxy was added to each quasar image before subtraction of a stellar PSF. The galaxy sizes were held fixed as they appear in Figure 5. In all the cases considered in this paper, the center of the putative host galaxy and the quasar were aligned. Host galaxies are more easily detected if they are not centered on the quasar (see Paper I).

The $\chi^2$ for the best PSF fit was calculated as a function of galaxy brightness. As the galaxy image was made progressively fainter, the $\chi^2$ dropped until the galaxy signal became lower than the background noise. When the computed $\chi^2$ equaled twice the asymptotic value, a clear break point in the $\chi^2$ vs brightness curves, we could easily detect the galaxy placed underneath the quasar. Visual inspection of the subtracted images showed that in some cases (in particular for the edge-on spiral, Figure 5b), a galaxy could be detected $\approx 1$ mag fainter than indicated by the $\chi^2$ test described above.

Figure 6 shows the typical dependence that is found for $\chi^2$ as a function of apparent magnitude of the putative host galaxy.

Table 3 gives, for all eight of the quasars analyzed so far, the limiting magnitude down to which each of the eight galaxies shown in Figure 5 could be detected when the galaxy was centered on the quasar image. The limiting magnitudes presented in Table 3 take account of the information obtained by visual inspection.

For a given quasar, the limits of detectability of the galaxy images spanned a range of $1.5 \pm 0.3$ mag (see Table 3). Table 1 presents (in the next-to-last column) the magnitude ranges at which the faintest host galaxies, fainter versions of Figure 5a to Figure 5h, could be detected. We averaged, for each quasar, the limiting magnitudes given in Table 3 for each of the eight galaxies shown in Figure 5 and used this average magnitude to compute a representative limiting magnitude of a detectable host galaxy. The last column of Table 1 gives for all eight quasars the average absolute magnitude at which the eight galaxies in Figure 5a-h were visible when placed underneath that quasar.

Averaging all 64 determinations of limiting detectability for the eight galaxies, Figure 5a-h, placed underneath the eight quasars, we find that the ensemble-average apparent magnitude at which host galaxies could be detected is 19.3, about 4.1 mag fainter than the average quasar magnitude. The average absolute magnitude limit is

$$M_{\text{F606W}}(\text{lim}) = -19.6, \qquad (1)$$

which is about one magnitude fainter than $L^*$ for field galaxies (cf. footnote a of Table 1).

In general, spirals are easier to detect than ellipticals, because spirals have more features. On average, the spirals could be detected to a limiting magnitude of 19.4 mag, i. e.,

$$M_{\text{F606W}}(\text{lim, spirals}) = -19.5. \qquad (2)$$

The two early-type galaxies, Figure 5d-e, could only be seen down to 18.3,

$$M_{\text{F606W}}(\text{lim, early type}) = -20.6, \qquad (3)$$

a magnitude brighter than the ensemble average. Typically, we can detect the interacting galaxies shown in Figure 5g,h down to 20.1, 0.8 mag fainter than the average of the other six galaxies shown in Figure 5.

Table 3: Limiting apparent magnitudes

| Galaxy | 0953+414 | 1116+215 | 1202+281 | 3C 273 | 1302−102 | 1307+085 | 1444+407 | 3C 323.1 |
|---|---|---|---|---|---|---|---|---|
| 5a | 19.7 | 19.1 | 19.7 | 17.7 | 19.4 | 19.5 | 20.3 | 20.4 |
| 5b | 20.5 | 19.8 | 20.1 | 18.0 | 20.0 | 20.0 | 20.4 | 20.5 |
| 5c | 18.7 | 18.8 | 18.8 | 17.0 | 18.7 | 18.5 | 19.7 | 19.8 |
| 5d | 18.1 | 18.9 | 18.8 | 17.0 | 18.4 | 18.3 | 19.6 | 19.0 |
| 5e | 18.8 | 18.1 | 18.1 | 18.1 | 18.1 | 18.1 | 18.1 | 18.1 |
| 5f | 20.5 | 19.0 | 20.0 | 18.0 | 19.5 | 19.4 | 20.4 | 20.2 |
| 5g | 20.1 | 20.0 | 20.3 | 18.8 | 21.2 | 20.5 | 21.3 | 21.3 |
| 5h | 19.7 | 19.0 | 19.8 | 18.0 | 20.2 | 20.0 | 20.0 | 20.6 |
| 5bx3 | 19.4 | 19.3 | 19.5 | 18.8 | 20.2 | 20.0 | 20.5 | 21.3 |
| 5fx3 | 19.7 | 19.0 | 19.7 | 18.5 | 19.9 | 19.7 | 20.8 | 20.8 |

Some previous authors have suggested that the host galaxies of quasars were large in size. In order to simulate large host galaxies, we also studied scaled versions, three times larger but with the same surface brightness profiles as in Figure 5, of the galaxies in Figure 5b and Figure 5f. The results for the two simulated larger galaxies are given in the last two rows of Table 3. On average, we could detect these large putative host galaxies to 19.8 mag, about 0.5 mag fainter than the ensemble average of the eight observed galaxies shown in Figure 5.

We are able to detect galaxies that are well-separated from the quasars down to about $24^{th}$ mag, approximately five magnitudes fainter than the limiting magnitudes listed for host galaxies in Table 3.

## 4. Fits to model galaxies

Many authors have determined the magnitudes of host galaxies by fitting the quasar images to a model consisting of a point source plus an analytic model of a featureless galaxy (an exponential disk or a de Vaucouleurs profile) (see, e.g., McLeod & Rieke 1994a,b; Dunlop et al. 1993; Véron-Cetty & Woltjer 1990; and references therein). This procedure assumes that the host galaxies are smooth and featureless (which is different from most of the HST-observed galaxies shown in Figure 5 of § 3), that the standard stellar PSF is a good match to the PSF of the nuclear region of the quasar, and that there are no bright companion galaxies.

We shall see in this section that, not surprisingly, we obtain limits on, or possible measurements of, host galaxy magnitudes that are about one magnitude less stringent when we fit the residual light to smooth galaxy models than when we determine ( as in § 3)) the limits of detectability of HST-observed spiral galaxies. This difference also reflects the approximately one magnitude difference in the detectability limits found in the previous section (see Equation 2 and Equation 3) between spirals and early-type galaxies.

We have fit combined models of a stellar nucleus plus a host galaxy with an exponential disk to each of the long-exposure quasar images. The quasar nucleus was approximated by the composite stellar PSF (Fig. 2). The general form of the assumed fit was

$$\text{Light}(x,y) = C_1 \times \text{PSF}(x,y) + C_2 \times \text{Host}(x,y; \text{length scale}) + \text{Sky Offset}, \qquad (4)$$

where the PSF and the host galaxy intensity, Host, were normalized to unity when integrated over all positions, (x,y). We used analytic functions, an exponential disk or a de Vaucouleurs profile, for the host galaxy surface brightness.

For many, but not all, of the tests we ran, the central intensity and position of the stellar PSF were fixed so as to best-eliminate the diffraction spikes (cf. Fig. 4). We varied the initial guesses for the amount of light in the stellar nucleus, $C_1$, over a wide range and also varied by large factors the initial guesses for the amount of light, $C_2$, in the host galaxy. In addition, we varied the characteristic length scale, the effective radius (for a de Vaucouleurs profile) or the scale length (for an exponential disk), between $1''$ and $3''$. Moreover, we did experiments in which the fit was optimized within different annuli; typically, the annulus used was $0.5''$ to $5''$, but we also used other annuli (such as $2.0''$ to $6.5''$).

The formal $\chi^2$ fits to the combined model of a smooth exponential disk plus stellar nucleus were generally poor; the typical $\chi^2$ per degree of freedom was 2 to 4, but in a number of cases the fit was much worse. Some of the mismatch was caused by the imperfect subtraction of the stellar PSF and is evident in Figure 4; we could not completely remove the diffraction spikes in the subtracted images of the quasars. In retrospect, it is perhaps not surprising that the residual light is not well described by a featureless, smooth analytic galaxy model.

The results for the best-fitting apparent and absolute magnitudes of the model host galaxies are shown in Table 4 for the special case in which the initial guess for the parameters was $C_1 = C_2 = 0.0$. In general, the fits with an exponential disk were rather stable to parameter variations, with a total range of only $\pm 0.2$ in the average absolute magnitude for the sample. The average absolute magnitude for the best-fitting exponential disks of the host galaxies is

$$< M_{\rm F606W}({\rm lim}) >_{8 \text{ quasars}} = -20.5 \pm 0.2 \text{ mag}, \qquad (5)$$

which is equal to the absolute magnitude of an $L^*$ galaxy. A similar range, $\pm 0.2$ mag, was found for the variations of most of the fits to the magnitudes of the assumed host galaxies in individual quasars, although occasionally differences as large as $\pm 0.5$ mag were found for individual model host galaxies.

The magnitudes given in Table 4 and in Equation 5 must be regarded with caution.

We have no proof that the residual light being fitted by the smooth galaxy models is in the assumed form of a smooth exponential disk. In fact, the fits to this form are, as mentioned above, poor. Moreover, if we start the $\chi^2$ optimization routine at a large negative value for the host galaxy intensity, i.e., $C_2$ is negative and large, we find host galaxies with negative fluxes. If we change the sign of the negative fluxes obtained in this way, they correspond to host galaxy magnitudes of about the same average brightness as found when we made an initial positive guess for the light in the host galaxy, i.e., they are consistent with Equation 5. The scale lengths for the cases with negative fluxes, however, are all very large, so large that the surface brightness of the galaxies is comparable to the noise in the sky. One indication of the size of the error in the fitting procedure may be the absolute magnitudes obtained by inverting the sign of the negative fluxes, which is of order of $M_V = -20.5$ mag.

We also find it somewhat suspicious that all of the exponential scale lengths cluster around the values $1.2 \pm 0.3''$. This clustering suggests that the fits could be representing some instrumental feature instead of an astronomical property.

The fits to a de Vaucouleurs profile were even less stable to variations of the initial parameters than were the fits to an exponential disk. The best-fitting effective radii were sometimes very large, comparable to the outermost radius considered in the fitting process. In a few of the cases in which a de Vaucouleurs profile was assumed, we found negative fluxes for the host galaxies even if the initial value of $C_2$ was taken to be zero. These cases had very small effective radii, typically of order $0.5''$, so that a large fraction of the light in the model host galaxy occurred in areas that were not part of the fitting region.

We conclude that there is some evidence that there is residual light not associated with the stellar nuclei of the quasars that has an average absolute magnitude of about $L^*$. However, this light is not well-described by a smooth galaxy model. In addition, changes in the initial guesses for the fitting parameters can sometimes produce negative galaxy fluxes. We conclude that our observations are not able to determine decisively whether smooth, featureless galaxies are present with visual luminosities about equal to $L^*$.

Observations with HST at different telescope roll angles are required in order to decide if the residual light that is fit in this section by smooth galaxy models is associated with the quasar or is somehow produced in the WFPC2.

Table 4: Total Magnitudes for a Model of a Point Source plus an Exponential Disk

| Quasar | $z$ | $m_{606W}$ exp. disk | $M_V$ exp. disk |
|---|---|---|---|
| PG 0953+414 | 0.239 | 19.5 | −19.9 |
| PG 1116+215 | 0.177 | 18.1 | −20.7 |
| PG 1202+281 | 0.165 | 18.7 | −19.8 |
| 3C 273 | 0.158 | 16.8 | −21.7 |
| PKS 1302−102 | 0.286 | 18.4 | −21.4 |
| PG 1307+085 | 0.155 | 19.0 | −19.4 |
| PG 1444+407 | 0.267 | 18.8 | −20.8 |
| 3C 323.1 | 0.266 | 19.1 | −20.5 |

## 5. Characteristics of Candidate Host Galaxies

Tests described in this paper (see § 8) and in Paper I show that we have detected diffuse emission in excess of the presumed stellar nuclear component (as modeled by the composite PSF shown in Figure 2) for at least some of the quasars we have studied. However, the available HST observations are insufficient to establish definitively whether we have detected residual light resulting from: 1) the faint host galaxies of quasars; 2) the difference between the composite PSF of the red star F141 (Fig. 2) and the PSF of the blue quasar nucleus; or 3) excess light from companion galaxies that are projected very close to the quasars (see, e.g., the close companions of PG 1202+281, PKS 1302−102, and 3C 323.1 in Figure 1 of Paper I and Figure 4 of this paper). In this section, we present crude estimates of the characteristics of candidate host galaxies for three quasars, based upon the assumption that all of the detected residual light is contained in galactic hosts.

For the three objects PG 1116+215, 3C 273, and PG 1444+407, we have detected residual emission that is plausibly attributed to host galaxies. In all three cases, the characteristic size, defined to be one-half the major axis of the diffuse emission, is about 4″.

Table 5 lists the estimated characteristics for the three candidate host galaxies. The columns contain the following information: the quasar name, the estimated aperture magnitude, the estimated magnitude determined by assuming that the host galaxy has an exponential disk profile, the absolute magnitude of the host galaxy computed from the aperture magnitude (or the exponential disk model) assuming that it has the same redshift as the quasar, the average limiting absolute magnitude at which the galaxies in Figure 5a-h could just be detected (cf. § 3.3), and comments regarding the galaxy.

The aperture magnitudes were determined by measuring the residual surface brightness in regions well-separated from the quasar (typically at radial distances of about 3″) and then multiplying by the approximate area over which the surface brightness extended. This procedure yields a reasonable estimate for the diffuse emission only if the host galaxy profile is flat and if the residual light is all due to a host galaxy. The exponential disk magnitudes were determined by finding the parameters that best-fit the quasar image assuming all of the light is produced by the composite PSF of Figure 2 plus an exponential disk galaxy (see discussion in § 4).

Radiation from the quasar will have a strong effect on the interstellar medium of the host galaxy (see, e.g., Begelman 1985 and Chang, Schiano, & Wolfe 1987). The radiation may strongly inhibit star formation near to the quasar. Therefore, it is not obvious whether an aperture magnitude calculated assuming a constant surface brightness or an exponential disk model with a surface brightness increasing toward the quasar will give a more accurate estimate for the total magnitude of a host galaxy. Moreover, our experiments with fitting model host galaxies to our quasar images suggest that the results are of uncertain significance.

In the following discussion, whenever a best-estimate for the magnitude of a host galaxy is required we use the average of the aperture and the exponential disk magnitudes listed in Table 5.

Table 5: Estimated Characteristics of Candidate Host Galaxies

| Quasar | $m_{606W}$ (apert. mag) | $m_{606W}$ (exp. disk) | $M_V$ | $\langle M_V \rangle_{\text{limit}}$ | Comments |
|---|---|---|---|---|---|
| PG 1116+215 | 19.0 | 18.1 | $-19.7(-20.7)$ | $-19.6$ | Partial rings? |
| 3C 273 | 18.0 | 16.8 | $-20.5(-21.7)$ | $-20.9$ | Off-center elliptical? |
| PG 1444+407 | 19.7 | 18.8 | $-20.0(-20.8)$ | $-19.6$ | Ring-like, bar? |

For PG 1116+215 (see Fig. 4b), there appears to be a faint (23.4 mag arcsec$^{-2}$), smooth ring-like structure with a radius of about 2″ surrounding almost half of PG 1116+215, plus an additional diffuse protrusion. If this nebulosity is real, it corresponds to a host galaxy with an aperture magnitude of $m_{F606W} \approx 19.0$ (0.1 mag brighter than the average limiting magnitude estimated for the eight HST-observed galaxies listed in Table 3), and a total diameter of $\sim 20$ kpc. We are somewhat suspicious of this candidate nebulosity because of its smooth, symmetric shape.

The image of the bright radio quasar 3C 273 shows evidence of an elliptical host in Figure 4d, which displays the residual image after subtraction of a best-fit point source. The candidate host galaxy is relatively bright, with an aperture magnitude of $m_{F606W} \approx 19.0$. However, the nebulosity does not appear to be centered on the quasar. If the host galaxy is symmetric, then the separation of the quasar and the center of the host galaxy is between 1″ and 2″ (2 to 4 kpc). We are also somewhat suspicious of this emission because of the relatively sharp appearance of the outer arc that forms the western boundary of the nebulosity.

The most conventional candidate we have found for a host galaxy is associated with PG 1444+407 and can be seen clearly in the residual image Figure 4g. This faint nebulosity has an approximate aperture magnitude of $m_{F606W} \approx 19.7$ and is centered on the quasar.

The estimated average absolute magnitudes of the candidate host galaxies are in all three cases close to the limiting absolute magnitudes at which host galaxies could be detected (as defined by the experiments discussed in § 3.3). The differences are (cf. Table 5): 0.6 mag (PG 1116+215), 0.2 mag (3C 273), and 0.8 mag (PG 1444+407). One would expect that test galaxies placed underneath quasars with real host galaxies would not be detectable if the apparent magnitudes of the test galaxies are of the order of, or significantly fainter than, the apparent magnitudes of the true host galaxies. However, it is always a source of concern if one finds that the quantities measured are close to the limits of detectability.

We also note (cf. Paper I) that for PG 0953+414 we see what may be an extended, low surface brightness feature ($\mu_V \sim$ 24 mag per square arcsecond, between $PA \approx 100°$ to $\sim PA \approx 180°$ ) at $\sim 3''$ to $\sim 5''$ from the quasar image.

## 6. Companion Galaxies

The unprocessed images of the quasar fields displayed in Figure 1 reveal a number of companion galaxies projected close to the quasars. Very close companions (separations $\lesssim 2''$) are shown more clearly in Figure 4, which presents the residual images after subtraction of a stellar PSF.

In order to evaluate the statistical significance of the small projected separations of the quasar-galaxy pairs, we decided to count the number of companion galaxies brighter than a specified limiting absolute magnitude that were found to have a metric separation from one of the quasars of less than or equal to some fixed distance. We choose *a priori* a limiting absolute magnitude of $M_V = -16.5$ (four magnitudes fainter than $L^*$) and a maximum separation of 25 kpc. All galaxies found around the quasars that satisfy these specifications are included in our complete sample of galaxy companions.

Table 6 lists for each quasar the number of companion galaxies that are at least as bright as $M_V = -16.5$ and that are projected within 25 kpc of the center of light of the quasar. The table gives the separations both in arc seconds and, in parentheses, in kpc. Similarly, the brightnesses of the galaxies are tabulated in both apparent and absolute magnitude.

How can we evaluate the statistical significance of the close pairs listed in Table 6? The ideal procedure would be to count galaxies in random areas that were also imaged with WFC2 and the same filter used here, F606W. We intend to do this when the full sample of 18 quasar fields is available. In the meantime, we can establish an *upper limit* to the random probability by counting galaxies in comparison regions imaged at the same time and with the same instrumental configuration as each of the quasar fields. We count galaxies in circular areas that are within 25 kpc (at the quasar redshift) of the quasar position shifted to Chip 2 and Chip 4. (As described earlier, see § 2, the quasar images are all close to the center of Chip 3 of the WFC2.) This procedure provides an upper limit to the random probability of finding galaxies of the specified minimum brightness within the given area because the centers of Chip 2 and Chip 4 are separated by only 80", about 170 kpc at $z = 0.2$, from the center of Chip 3. The separation of the centers of Chip 2 and Chip 3 (or Chip 4 and Chip 3) is, for objects at $z = 0.2$, rather similar to the typical core radius ($\sim 125$ kpc, see Bahcall 1977) of a rich Abell cluster of galaxies. Therefore, the comparison regions in Chip 2 and Chip 4 may contain an enhanced number of galaxies if the quasar-galaxy correlation function is large at low redshifts. The areas inspected in the comparison regions are slightly larger than the accessible areas around the quasars, since companion galaxies are difficult to observe on our images if they are projected less than 1" from the quasar centers.

Table 6: Galaxy companions brighter than $M_{\text{F606W}} = -16.5$ within 25 kpc of the quasar

| Quasar | Number of Companions | Distances " | Distances kpc | Magnitudes $m$ | Magnitudes $M$ |
|---|---|---|---|---|---|
| PG 0953+414 | 1 | 8.2 | 19.6 | 22.9 | $-16.5$ |
| PG 1116+215 | 1 | 12.6 | 24.3 | 19.5 | $-19.2$ |
| 1202+281 | 2 | 5.2 | 9.5 | 19.3 | $-19.3$ |
|  |  | 9.6 | 17.7 | 21.9 | $-16.7$ |
| 3C 273 | 0 | – | – | – | – |
| PKS 1302−102 | 2 | 1.1 | 3.0 | 20.6 | $-19.2$ |
|  |  | 2.3 | 6.2 | 21.9 | $-17.8$ |
| PG 1307+085 | 0 | – | – | – | – |
| PG 1444+407 | 0 | – | – | – | – |
| 3C 323.1 | 1 | 2.7 | 6.9 | 20.8 | $-18.8$ |

In the 16 fields of Chip 2 and Chip 4 that accompany the eight quasar images studied here, there are 6 galaxy companions that would have satisfied the *a priori* criteria for inclusion in our complete sample of close companions had they been found at the same coordinate positions in Chip 3.

We have also counted galaxies close to four other quasars that are in our complete sample of 18 quasars and for which data became available after the initial drafts of the present paper were finished. These new objects are PKS 2135−14 ($z = 0.200$), PKS 2349−014 ($z = 0.173$), PHL 909 ($z = 0.171$), and NAB 0205+02 ($z = 0.155$). There are five additional close galaxy companions (two for PKS 2349−014 and one each for the other three new quasars) in the four new quasar fields; there are no new companion galaxies in the eight new comparison regions (of Chip 2 and Chip 4).

If we limit ourselves to the eight quasar fields studied in the present paper, then the average number of companion galaxies in Chip 2 or Chip 4 within the specified parameters is 3/16 or 0.375. The Poisson probability of finding seven or more projected close galaxy companions as listed in Table 6 is

$$P(\geq 7; 8; 25 \text{ kpc}, M \leq -16.5) \leq 0.02. \qquad (6a)$$

If we consider all twelve quasar fields imaged so far by HST, then the average number of companion galaxies in Chip 2 or Chip 4 is 6/24 or 0.250. The Poisson probability of finding at least the twelve projected companions observed close to the twelve quasars is

$$P(\geq 12; 12; 25 \text{ kpc}, M \leq -16.5) \leq 0.0001. \qquad (6b)$$

We conclude that the presence of the projected close companions seen in Figure 1 and Figure 4 is statistically significant.

The very close companions, separations $\lesssim 3''$, are difficult to observe from the ground. There are three such very close companions of the eight quasars considered here that are brighter than our absolute magnitude limit of $M_V = -16.5$ and none within $3''$ of the 16 equivalent quasar positions on Chips 2 and 3. If the three very close companions are not physically associated with their neighboring quasars but instead are a chance superposition, then the *post facto* Poisson probability for finding no companions within $3''$ for the 16 comparison regions is 0.0025. There is one additional very close companion brighter than $M_V = -16.5$ among the four new quasars (projected close to PKS 2349−014), but no galaxies within 3″ of the equivalent quasar positions on Chip 2 and Chip 4. The *post facto* Poisson probability of finding a total of four very close companions (closer than 3″) among the 12 quasar fields and zero very close companions in the 24 comparison regions is less than 0.01%, but *post facto* probabilities are often very small.

The situation for PKS 1302−102 is especially remarkable and is shown in Figure 7. Two companions of this quasar are seen clearly in Figure 7; they are projected at separations of only 1.1″ (3 kpc) and 2.3″ (6 kpc) from the quasar location. They lie at projected distances that are well within what would be the volume enclosed by a typical $L^*$ host galaxy. Hutchings and Neff(1990) report the presence of an extremely luminous host galaxy, $M_V = -23.7$, which we do not detect. Hutchings et al. (1994) did detect the companion at 2.3″; it appeared to be an "extended knot" with the combined data from the original WFC and CFHT HRCam.

## 7. Brightness Limits on McLeod-Rieke Host Galaxies

Recently, McLeod & Rieke (1994a,b) have performed a very careful ground-based study in the $H$−band (1.6$\mu$) searching for host galaxies of quasars; all eight of the quasars discussed in this paper are in their sample. McLeod & Rieke (1994b) report detecting bright host galaxies in at least 23 of 26 high-luminosity, nearby quasars ($M_B < -23.1$ mag and $z \leq 0.3$). They conclude that the high luminosity quasars have relatively bright host galaxies, typically twice the luminosity of an $L^*$ (Schechter) galaxy. Their results are similar to those of a number of other workers who have used ground-based data in different wavelengths (see, e.g., Hutchings & Neff 1992; Véron-Cetty & Woltjer 1990 and other references cited in these papers and in McLeod & Rieke 1994a,b).

Since the McLeod and Rieke observations were made in $H$ band and our observations are carried out in essentially the $V$ band, we cannot compare our results directly with theirs. However, we can compute the average absolute magnitude in $V$ to which we would be sensitive for host galaxies like those reported by McLeod and Rieke. By comparing our limit on $< M_V >$ with their reported $< M_H >$, we can infer a minimum value of $V - H$ for the two sets of results to be consistent.

We have created simulated host galaxies which have the same exponential scale lengths as the host galaxies described by McLeod & Rieke (1994b) in order to determine our ability to detect hosts of the type they described. The simulated galaxies we used are more difficult to detect than real galaxies with the same scale lengths. The simulated galaxies have no particular features that the computer can notice; they are smooth by construction.

All of the host galaxies reported by McLeod & Rieke (1994b) for the quasars shown in Table 7 have apparent exponential scale lengths that are of order $1.25''\pm 0.5''$, i.e., typically of order 2.5 kpc at the redshifts of the quasars. The seeing in their observations was also about $1.25''\pm 0.25''$ (see McLeod & Rieke 1994b ). The similarity between the quoted seeing and the model scale lengths may be an indication that some caution might be required in estimating the systematic uncertainties. McLeod and Rieke did not publish their scale-lengths because they used the exponential galaxy profiles as fitting-functions, without claiming that the profiles represented real galaxies. We are grateful to Dr. K. McLeod for generously making available her unpublished values for the scale lengths.

We created two simulated, face-on, smooth galaxies with exponential scale lengths of $1.0''$ and $1.5''$. We then performed tests identical to those described in Section 3.3 to determine how faint such galaxies could be and escape detection in our tests.

The results are shown in Table 7. On average, we are sensitive to the simulated host galaxies to $V = 18.7$, which corresponds, for the redshifts of the eight quasars we investigate, to $M_V = -20.4$. Thus we are able to detect the smooth featureless simulated galaxies to 0.1 mag fainter than $L^*$. Requiring our results to be consistent with those of McLeod & Rieke (1994b), would imply that the host galaxies of the eight quasars we studied are at least one magnitude redder in $V - H$ than ordinary field galaxies.

## 8. WHAT HAVE WE DONE WRONG?

We are painfully aware that our initial results appear to be in conflict with suggested detections of relatively bright quasars by previous authors. Therefore, we have performed a number of tests of our analysis procedures to see if we have made any errors that might explain the difference between our results and previously published results. Unfortunately, we have not discovered an explanation. We list below some of the tests that we did perform.

We first compare in § 8.1 the ground-based PSFs used in previous work on host galaxies of quasars with the point spread function measured for our HST observations. Next, we investigate in § 8.2 the possible uncertainties that result from errors in the sky subtraction and then determine in § 8.3 the likely uncertainties that are introduced by using a stellar point spread function that might be somewhat different (because of spatial, temporal, or color changes in the PSF) than the PSF of the nuclear region of the quasar. Next we determine in § 4 the best-fit magnitudes for the host galaxy if require that all of the residual light, after subtraction of a stellar PSF from the quasar, be attributed to a host galaxy with an exponential disk or a de Vaucouleurs profile. The intensity of the best-fitting stellar PSF may be determined in at least two extreme ways, either by minimizing the residual light by optimally matching the diffraction spikes or by azimuthally averaging the light in the quasar images. In § 8.4, we show that both methods yield essentially the same results. Finally, we construct in § 8.5 pseudo-AGNs and show that we can detect the presence of host galaxies approximately 4.5 mag fainter than the simulated quasars.

Table 7: Detectability of Smooth, Face-On Exponential Disk Galaxies

| Quasar | Scale Length | |
|---|---|---|
| | 1.0″ | 1.5″ |
| PG 0953+414 | 18.5 | 18.8 |
| PG 1116+215 | 18.0 | 18.7 |
| PG 1202+281 | 19.1 | 19.5 |
| 3C 273 | 17.0 | 17.7 |
| PKS 1302−102 | 18.5 | 18.9 |
| PG 1307+085 | 18.0 | 18.7 |
| PG 1444+407 | 18.9 | 19.5 |
| 3C 323.1 | 19.2 | 19.5 |

## 8.1. Comparison of Ground-based and HST PSFs

Can HST detect faint diffuse emission near a bright stellar source more effectively than previous ground-based observations? The measurement of extended scattered light in the WFPC2 instrument (Krist and Burrows 1994) requires us to give a quantitative answer to this question.

Figure 8 shows a direct comparison of the PSF produced by the WFPC2 (see Figure 3) with the seeing profile in an excellent ground-based observation. Both curves present azimuthal averages. The HST PSF is shown in two parts; points inside of $1''$ are from a theoretical model by Burrows (1994b) and points beyond $1''$ are from our composite PSF (Figure 3). The ground-based seeing profile is from observations taken with the Hale telescope; the FWHM of the profile is $0.87''$ and the half-energy diameter is $1.01''$.

At a radius of $1''$, the HST PSF has an advantage of about a factor of ten over the ground-based HALE PSF. This advantage decreases somewhat at large radii because of scattering in the WFPC2 (Krist and Burrows 1994).

How good is the seeing used in the previously-published studies of the host galaxies of quasars? In the earliest papers on this subject (see references in § 1), the seeing conditions were often not reported, but in recent years most authors have reported the FWHM of the seeing either quantitatively or qualitatively. Table 8 gives the FWHM of the PSF as reported in the previously-published papers in which we found a definite statement about the seeing. The $0.87''$ Hale Telescope PSF shown in Figure 8 is better than nearly all of the PSFs used in previous ground-based work.

We conclude that the WFC2 PSF of the Hubble Space Telescope provides an important advantage over previous ground-based studies for projects in which one tries to detect faint diffuse emission near a bright stellar source.

## 8.2. Sky Subtraction

Oversubtracting the sky brightness would result in removing signal from a possible host galaxy. Sky oversubtraction is the astronomical equivalent of throwing out the baby with the bath water.

We have therefore measured the sky background over the fields of each of our CCD images. There are detectable (non-Poisson) variations in the flat-fielded sky intensity at the several percent level; the local deviations are smaller in the centers of the fields where quasar images are located than at the edges. (The sky levels in the corners of some fields are as much as 20% lower than at the centers of the fields.)

Table 8: The seeing conditions (FWHM) reported by different observers of quasar host galaxies

| Observers | Reported Seeing (FWHM) |
|---|---|
| Wyckoff, Wehinger, & Gehren (1981) | 1.1″–1.6″ |
| Hutchings et al. (1982) | ∼ 1″ |
| Malkan, Margon, & Chanan (1984) | 1.5″–2.0″ |
| Smith et al. (1986) | 0.9″–1.8″ |
| Hutchings (1987) | 0.9″–1.3″ |
| Hutchings, Janson, & Neff (1989) | 0.7″–1.3″ |
| Romanishen & Hintzen (1989) | 1.1″–2.4″ |
| Véron-Cetty & Woltjer (1990) | 0.9″–1.2″, 78% |
|  | 1.3″–1.4″, 22% |
| Hutchings & Neff (1992) | ∼ 0.5″ |
| Dunlop et al. (1993) | "complex" (0.6″ pixels) |
| McLeod & Rieke (1994b) | 1.25″ ± 0.25″ |

The measured sky brightness for the eight quasar images we have studied range (probably due to different amounts of earth-shine) from 22.2 mag arcsec$^{-2}$ (for 3C 273) to 23.0 mag arcsec$^{-2}$ (for PG 1444+407). Given the measured small fluctuations in the sky brightness across a CCD in a given exposure, it is extremely unlikely that the estimated sky levels are incorrect by as much as 10%. Even a ten percent error in the sky brightness would not affect significantly our inferred limits (Table 3) on the typical host galaxy magnitudes unless the host galaxies were as faint as $m_{F606W} = 21$, which is much fainter than the reported magnitudes of host galaxies determined from ground-based measurements (see Table 7) and also fainter than our limiting sensitivity (see Table 3).

We conclude that uncertainties in determining the sky brightness are probably not important in detecting host galaxies at the magnitudes discussed in this paper.

### 8.3. Subtracting One Stellar PSF From Another Stellar PSF

How accurate is the PSF subtraction that we perform to detect the host galaxies? Could imperfect PSF subtraction cause us to not detect luminous host galaxies of the quasars? Are the residual diffuse features seen in Figure 4 associated with the quasar or are they the result of imperfect stellar PSF subtraction?

In order to answer these questions, we selected a representative set of nine saturated stellar images that are contained in different WFC2 HST fields. The combined PSF of the standard star F141 in M67 (see Fig. 2) was subtracted from the image of each one of the other nine stars. The subtracted images are displayed in Figure 9.

The first six panels, $a - e$, of Figure 9 show the results obtained when the standard PSF of Figure 2 was subtracted from one of the other saturated stellar images, both the standard PSF and the other saturated stellar image having being obtained with the F606W filter. The last three panels of Figure 9 show the residual images found when the standard PSF was subtracted from stellar images observed through the F814W, F658N, and F439W filters, respectively.

The brightest residual diffuse emission is found in Figure 9d. This is also the case for which the original star had the largest saturated area of any of the test stars, considerably more than even in our longest exposures for the quasars. Moreover, the residual diffuse emission does not have the appearance of a galaxy. Among other things, the image contains a second, fainter star within 2″ of the center of the bright star. Also, part of the residual emission is contributed by what is apparently an optical ghost (see Paper I for the discussion of an another apparent doughnut-shaped optical ghost in the image of PG 1307+085). The relatively poor subtraction shown in Figure 9g also seems to be caused because the field star is highly saturated, again considerably more than our quasar images.

The residual surface emission shown in Figure 9 is probably an upper limit to what might be produced by inaccuracies in the stellar PSF subtraction from the actual quasar images.

The reason it is an upper limit is that the standard stellar PSF, Figure 2, was obtained with the star observed through the same filter as the quasar and located at the same position as the quasar image; we purposely mismatched PSFs in order to produce Figure 9.

We can estimate quantitatively the equivalent magnitudes of host galaxies that are produced by the mismatch between stellar PSFs, scaling the stars used to produce Figure 9 to the brightness levels that they would have if they were as bright as our average quasars, i.e., $m_{F606W} = 15.2$ mag. First, we calculate for each star the equivalent exposure time, $t_{equiv}$, that a 15.2 mag star would have been exposed in order to produce the number of photons counted for that star in the original image (before subtracting the standard PSF). Then we fit a model of the form given in Equation 4 to the stellar light and determine the total number, $N_{host}$, of photons that are in the component assigned to the host galaxy. Finally, we calculate the magnitude of the host galaxy using $t_{equiv}$ and $N_{host}$. When exponential disk models were fit to the HST stars, four of the pseudo-galaxies had positive counts and five had negative counts. The scale lengths were typically 20-40 pixels, and the pseudo-galaxies with positive flux had an average equivalent F606W magnitude of

$$< m_{F606W}(\text{mismatch}) >_{\text{exp. disk}} = 21.2, \tag{7}$$

and a median value of 20.8 mag. If one changes the sign of the negative flux pseudo-galaxies and assigns a magnitude to this value, the average brightnesses of these five objects is 20.9. If we estimate instead all of the light not in a point source using aperture magnitudes, we find considerably brighter residuals,

$$< m_{F606W}(\text{mismatch}) >_{\text{aper. mag}} = 19.8, \tag{8}$$

and a median value of 20.4 mag.

Even though the match between the standard and the field star PSFs is, by choice of the stellar images, not perfect, the subtraction is sufficiently accurate that there does not appear to be a significant danger that we have failed to observe luminous host galaxies ($m_{F606W} < 20.0$) because of errors in the subtraction.

### 8.4. Diffractions Spikes Versus Azimuthal Average

With ground-based observations, it is customary to determine the intensity of the stellar image to be subtracted from the quasar image by azimuthally averaging the light intensity near to the quasar and then requiring that azimuthal average to match as well as possible the PSF for a point source. This procedure has the disadvantage that close companion galaxies (some of which are visible on HST images; see Fig. 1, Fig. 4, or Fig. 7) are not properly taken into account. We are therefore led to investigate whether the azimuthal average that is usually employed in ground-based investigations might be a major source of the discrepancy between our results and some of the previous studies. In this subsection, we answer the following limited question: For the HST data, do we obtain significantly different estimates for the possible host galaxy magnitudes if we use an azimuthally-averaged radial surface brightness profile rather than concentrating on matching as well as possible the diffraction spikes? Our previous discussion in § 3 and § 5 was based upon an approximately optimal matching of diffraction spikes in the point source PSF and the quasar image.

In order to answer the question posed here, we calculated in two ways the apparent magnitudes of host galaxies with exponential disks that, together with the composite point source PSF, best fit (in a $\chi^2$ sense) the quasar images. We determined the central intensity of the best-fit stellar PSF plus exponential disk by: 1) minimizing the residuals from the subtraction using all of the emission in annuli centered on the quasar images; and 2) first minimizing the residuals from the quasar minus point source subtraction by considering only the diffraction spikes. The two methods gave consistent results: in no case was a candidate host galaxies seen when the subtraction was done one way but not seen when the subtraction was done the other way. In addition, the magnitudes found for the exponential disk galaxies were about the same using the two different techniques. On average, the magnitudes obtained when the diffraction spikes were matched before the exponential disks were fit to the residuals led to only slightly brighter "host galaxies". Specifically,

$$< m(\text{azimuthal av.}) - m(\text{spikes}) > \ = \ (0.17 \pm 0.19) \text{ mag}. \qquad (9)$$

We conclude that the difference in subtraction procedures of concentrating on either the diffraction spikes or an azimuthal average does not cause us to overlook luminous host galaxies.

### 8.5. Host Galaxies of Pseudo-AGNs

We have performed an overall test of our techniques for finding host galaxies by creating pseudo-AGNs within observed or simulated host galaxies and then applying our standard software to demonstrate that we can detect the host galaxies. The pseudo-AGNs were produced by adding host galaxies to a saturated image of a 15.5 mag bright star located 13.8″ from the center of Chip 2 of the WFC2; this star was also used in producing Figure 9f. The stellar PSF that was used for the AGN nucleus was chosen to be very different from the composite PSF shown in Figure 2 and, in fact, the image of the pseudo AGN nucleus even has a small doughnut-shaped ghost image that is clearly visible in Figure 9f. We are therefore making it purposely difficult to detect the faint galaxy underneath the AGN nucleus by creating a poor match between the nuclear PSF and the composite standard PSF. The galaxy images that were used to make the simulated AGNs included HST-observed galaxies from Figure 5, galaxies observed with ground-based telescopes and adapted to HST conditions, and simulated (smooth) galaxies.

Each host galaxy was given a sequence of successively fainter magnitudes, usually beginning with 1.5 mag to 2.0 mag fainter than the pseudo-nucleus. Just as described in Section 3 for the analysis of the quasar images, we subtracted a best-fit standard PSF (Fig. 2) for each case, and then analyzed the results with the same software as was used in studying the quasars. We also inspected visually the subtracted images, just as we did for the quasars. The results were similar to what was obtained in Section 3 for the quasar images plus the HST-observed galaxies from Figure 5. Moreover, the $\chi^2$ distribution as a function of the faintness of the host galaxy had the same form as for the real data (cf. Fig. 6).

Figure 10 shows the images obtained when a simulated, nearly edge-on exponential disk galaxy (exponential scale length of 1.5″, approximately 3 kpc at $z = 0.2$) was added to the stellar source. The galaxy looks much like the HST galaxies we have observed except that it is smoother. In Figure 10a, the galaxy is 1.7 mag fainter than the quasar nucleus (a 15.5 mag stellar image). The galaxy is 0.5 mag fainter in each successive panel. The limit of detectability is 19.7 mag, which is the apparent magnitude of the galaxy in Figure 10d. If the quasar nucleus in Figure 10 had the same absolute magnitude, $M_V = -23.9$, as the average quasar in our sample, the we could detect the host galaxy of the simulated AGN down to approximately $M_V = -19.7$.

We conclude that our technique is capable of discovering moderately bright to luminous host galaxies of quasars at small redshifts.

## 9. SUMMARY AND DISCUSSION

We have studied the environments of eight of the intrinsically most luminous quasars ($M_V \leq -22.9$) that are found at small redshifts ($z \leq 0.3$). The principal results of our analysis are summarized below.

(1) Separate companion galaxies are found projected so close to the quasar location that they fall well within the volume that would be occupied by an $L^*$ host galaxy (see § 6). For example, PKS 1302−102 has two companion galaxies clearly visible in Figure 7 that lie at projected distances of 3 kpc and 6 kpc, respectively, from the center-of-light of the quasar emission. Dynamical friction will cause these very close companions to spiral into the quasar in a time short compared to the Hubble time if the companions have typical galaxy masses.

We have counted galaxy companions within a projected distance of 25 kpc of the quasar position that are brighter than $M_V = -16.5$ for all eight of the quasars discussed in this paper and for four additional quasars for which data were obtained after this paper was almost complete. There are (see § 6) seven (twelve ) companion galaxies found near the eight (twelve) quasars. By counting galaxies in comparison fields, we have established (cf. Equation 6a and 6b) that most of the observed close galaxy companions are physically associated with the quasars.

These statistical results provide strong evidence for the view (see references in § 1) that one important ingredient in the luminous quasar phenomenon is the presence of galaxy companions. The very close companions may be in the process of providing fresh fuel by spiraling into the quasar.

It will be of great interest, but difficult, to obtain redshifts for the companion galaxies that are listed in Table 6. The cosmological nature of the quasar redshifts, assumed in this paper, plus the statistical association between quasars and companion galaxies that is described above together imply that the redshifts of the companion galaxies will be similar to the quasar redshifts.

(2) For the representative set of HST-observed galaxies shown in Figure 5, we could have detected host galaxies of the quasars if they had an average absolute magnitude as bright as $M_{\rm F606W}({\rm lim}) = -19.6$ mag, i.e., to a limiting magnitude that is one magnitude fainter than the characteristic (Schechter) magnitude $L^*$ for field galaxies. This average limiting magnitude was established in § 3.3 by placing a sequence of progressively fainter versions of each of the galaxies shown in Figure 5 underneath each of the quasar images and analyzing the residual emission that is left over after a point source PSF was subtracted from the quasar plus added-galaxy.

(3) We have fit smooth, featureless galaxy models (exponential disks or de Vaucouleurs profiles) to the residual light after a best-fitting stellar PSF was removed from the quasar images. The fits are generally poor, but they do reveal (§ 4) evidence for possible host galaxy light that, on the average, is about equal to $L^*$. This is also about the limit of detectability found in § 3.3 for early-type HST-observed galaxies.

(4) We have found plausible candidate host galaxies for three of the eight quasars we have studied, namely, for PG 1116+215, 3C 273, and PG 1444+407. However, these candidate host galaxies are not remarkably luminous; they have an average $V$-band brightness of about $L^*$. In fact, we are rather suspicious of the candidate host galaxies that we have found because they are all rather close to the limits of our range of detectability (typically within 0.5 mag of the estimated faintest detectable host galaxy, see Table 5) and have, in two of the three cases, properties that are unexpected for host galaxies (see § 5). HST observations at a different telescope roll angles of these three quasars would provide crucial tests for the reality of the candidate host galaxies. If the residual emission tentatively ascribed to host galaxies is real, it will have a fixed shape and position on the sky, independent of telescope orientation.

(5) Our results on the absence of luminous host galaxies are surprising given the generally accepted view that quasars reside in host galaxies with the most luminous quasars residing in the most luminous galaxies. It is difficult to compare our results directly with previous ground-based observations because of the higher angular resolution available in the HST images and, in many cases, because of the different photometric bands (or spectra) that were used. It is not clear how the presence of close companions (revealed by HST images) would affect measurements from the ground of the azimuthally-averaged residual light distributions remaining after a point-source PSF is subtracted.

In this paper, we have concentrated on analyzing carefully the HST data and on providing new quantitative results. We have generally avoided speculating on whether our data are consistent with previous ground-based observations. We believe that the observers who made the ground-based observations are best able to decide on the extent of consistency or lack of consistency between our results and theirs and to make new observations to test our results and theirs using the information provided by the HST images. We will do everything we can to facilitate the comparison of our results with previous and future ground-based results.

To illustrate some of the considerations that should be taken into account in a comprehensive study, we have compared the HST results with several ground-based observations for which previous observers have published quantitative results for quasars in the sample studied in the present paper. For example, we have made a global comparison of our analysis with the recent results of McLeod & Rieke (1994b), who have reported finding bright host galaxies in $H$-band images for all eight of the quasars studied in the present paper. In order for our results to be consistent with the McLeod-Rieke conclusions, the host galaxies of luminous quasars must, on the average, be at least one magnitude redder in $V - H$ than for normal field galaxies. Hutchings et al. (1989), on the other hand, have reported that the host galaxy of PG 0953+414 has $M_B = -20.9$, which would be consistent with our upper limit if the host galaxy has an extremely blue color, i.e., $B - V < -1.0$ mag.

Hutchings and Neff(1990) report, based upon ground-based observations, the presence of an extremely luminous host galaxy for PG 1307+085; they give an absolute magnitude of $M = -23.7$, much brighter than the brightest galaxies of rich clusters. Although their observations were taken in $R$ and ours were taken in $V$, we have not been able to think of a plausible explanation of why our upper limits for the host galaxy light are so much less than their detection. Our average upper limit for host galaxies like the HST-observed galaxies in Figure 5 is $M_V = -20.1$ (see Table 1) and the exponential disk fit to the total residual light is $M_V = -21.4$ (see Table 4).

Véron-Cetty & Woltjer (1990) have suggested that the apparent magnitudes of host galaxies should be measured in a fixed metric annulus that is well removed from the quasar nucleus. They propose an annular region of 12.5 kpc to 25.0 kpc for $\Omega_0 = 0.0$ and $H_0 = 50$ km s$^{-1}$Mpc$^{-1}$. We have only one object in common with Véron-Cetty and Woltjer, PKS 1302−102. The annular region specified by Véron-Cetty and Woltjer is between 2.175″ and 4.35″ for PKS 1302−102, which includes contributions from the two faint companion galaxies shown in Figure 7. The $V$ magnitude that Véron-Cetty & Woltjer (1990) obtain after transforming their $i$-magnitude measurement for this annulus is $V = (19.46 \pm 0.2)$ mag. We obtain $m_{F606W} = (19.6 \pm 0.2)$, in good agreement with their result. However, Véron-Cetty & Woltjer (1990) estimate that the host galaxy has a total $M_V = -23.7$ for $\Omega_0 = 0.0$ and $H_0 = 50$ km s$^{-1}$Mpc$^{-1}$, which corresponds to $M_V = -22.0$ for the cosmological parameters used in this paper. The host galaxy reported by Véron-Cetty and Woltjer for PKS 1302−102 is as bright as the brightest galaxy of a rich cluster. For the same quasar, we are only able to set an average upper limit to the brightness of a host galaxy of $M_V = -20.1$ (see Table 1) if the hosts are like the HST sample shown in Figure 5. On the other hand, we obtain host galaxy luminosities in the range $M_V = -21$ to $M_V = -22$ if we fit the residual light to an exponential disk or a de Vaucouleurs profile and try various initial guesses for the model parameters. Because of difficulties in interpreting the results of fits to smooth model galaxies (see § 4), and because of the presence of close companion galaxies revealed by HST (see Figure 7 ), we are unable to reach any definitive conclusion in this case.

(6) Most of the work in preparing this paper was devoted to trying to answer the question: What have we done wrong? The longest section in this paper, § 8, describes a series of tests of our methods of analysis. These tests include: (1) measurements of the systematic errors due to sky subtraction; (2) a determination of the likely errors due to using a stellar PSF that is somewhat different from the PSF of the AGN nucleus; (3) adjusting the central intensity of the stellar PSF to either the diffraction spikes or the light within an annular region; and (4) constructing and analyzing simulated AGNs whose properties are known. Our analysis procedures successfully pass all of these tests. It is possible that some of the techniques we have described in § 8 for testing the validity of our methods of analysis may also be useful in connection with ground-based observations.

(7) Although our results are not consistent with the most luminous quasars residing in the most luminous known galaxies, our results are consistent with luminous quasars occurring in galaxies with a Schechter luminosity function that is cutoff at a moderate luminosity (Fall 1994). We have evaluated the probability that we would have obtained the eight brightness upper limits given in the last column of Table 1 (or more stringent limits), if the host galaxies have a Schechter luminosity function (cf. Efstathiou, Ellis, & Peterson 1988) that is cut-off on the low-luminosity side at $0.1L^*$ (or $0.01L^*$). The probability of obtaining a result as stringent as we have found is somewhat unlikely, 0.03, if the lower-cutoff in the luminosity function is $0.1L^*$(average host galaxy luminosity of $0.5L^*$), but is as large as 0.29 if the lower-cutoff is $0.01L^*$(average host galaxy luminosity of $0.2L^*$). The probabilities are even larger if we use the weaker limits that are obtained in § 4 for smooth featureless galaxies. In this case, we can approximate all of the upper limits by $L^*$. The probability of finding eight upper limits equal to $L^*$ (or fainter) is 0.4 (0.7) if $\epsilon = 0.1(0.01)$.

We are grateful to R. Blandford, C. Burrows, M. Fall, R. Gilliland, P. Goldreich, A. Gould, J. Gunn, J. Holtzman, B. Jannuzi, J. MacKenty, K. McLeod, J. Ostriker, D. Osterbrock, S. Phinney, G. Rieke, M. Rees, and M. Strauss for valuable discussions, comments, and suggestions. D. Saxe provided expert help with the figures. We would like to thank Digital Equipment Corporation for providing the DEC4000 AXP model 610 system used for the computationally-intensive parts of this project. This work was supported in part by NASA contract NAG5-1618 and grant number GO-2424.01 from the Space Telescope Science Institute, which is operated by the Association of Universities for Research in Astronomy, Incorporated, under NASA contract NAS5-26555.

# REFERENCES

Bahcall, J. N. 1986, ARA&A, 24, 577

Bahcall, J. N., Flynn, C., Gould, A., & Kirhakos, S. 1994, ApJ, L51

Bahcall, J. N., Kirhakos, S., & Schneider, D. P. 1994, ApJ, 435, L11

Bahcall, J. N., Schmidt, M., & Gunn, J. E. 1969, ApJ, L77.

Bahcall, N. A. 1977, ARA&A, 15, 505

Bahcall, N. A., & Bahcall, J. N. PASP, 82, 721 (1970).

Begelman, M. C. 1985, ApJ, 297, 492

Boroson, T. A., Persson, S. E., & Oke, J. B. 1985, ApJ, 293, 120

Burbidge, G. R., & O'Dell, S. L. 1973, ApJ, 182, L47.

Burrows, C. J. 1994a, Hubble Space Telescope Wide Field and Planetary Camera 2 Instrument Handbook, Version 2.0 (Baltimore: STScI)

Burrows, C. J. 1994b, private communication

Chang, C. A., Schiano, A. V. R., & Wolfe, A. M. 1987, ApJ, 322, 180

Chevalier, C., & Ilovaisky, S. A. 1991, A&A, 90, 225

Dunlop, J. S., Taylor, G. L., Hughes, D. H., & Robson, E. I. 1993, MNRAS, 264, 455

Eggen, O. J., & Sandage, A. R. 1964, ApJ, 140, 130

Efstathiou, G., Ellis, R. S., & Peterson, B. A. 1988, MNRAS, 232, 431

Fall, M. 1994, private communication

Gehren, T., Fried, J., Wehinger, P. A., & Wyckoff, S. 1984, ApJ, 278, 11

Gilliland, R. L. et al. 1991, AJ, 101, 541

Green, R. F., & Yee, H. K. 1984, ApJS, 54, 495

Gunn, J. E. 1971, ApJ, L113.

Heckman, T. M., Bothum, G. D., Balick, E., & Smith, E. P. 1984, AJ, 89, 958

Hoessel, J. G., & Schneider, D. P. 1985, AJ, 90, 1648

Huchra, J., & Burg, R. 1992, ApJ, 393, 90

Hutchings, J. B. 1987, ApJ, 320, 122

Hutchings, J. B., Crampton, D., Campbell, B., Gower, A. C., & Morris, S. C. 1982, ApJ, 262, 48

Hutchings, J. B., et al. 1994, ApJ, 249, L1

Hutchings, J. B., Janson, T., & Neff, S. G. 1989, ApJ, 342, 660

Hutchings, J. B., & Neff, S. G. 1990, AJ, 99, 1715.

Hutchings, J. B., & Neff, S. G. 1992, AJ, 104, 1

Hutchings, J. B., Morris, S. C., Gower, A. C., & Lister, M. L. 1994, PASP(to be published)

Kellermann, K. I., Sramek, R., Schmidt, M., Shaffer, D. B., & Green, R. 1989, AJ, 98, 1195

Kirhakos, S., Sargent, W. L. W., Schneider, D. P., Bahcall, J. N., Jannuzi, B. T., Maoz, D., & Small, T. A. 1994, PASP, 106, 646

Kirshner, R. F., Oemler, A., Schechter, P. L., & Shectman, S. A. 1983, AJ, 88, 1285

Krist, J., & Burrows, C. 1994, WFPC2 Instrument Science Report 94-01 (Baltimore: STScI)

Kristian, J. 1973, ApJ, 179, L61

Malkan, M. A. 1984, ApJ, 287, 555

Malkan, M. A., Margon, B., & Chanan, G. A. 1984, ApJ, 280, 66

McLeod, K. K., & Rieke, G. H. 1994a, ApJ, 420, 58

McLeod, K. K., & Rieke, G. H. 1994b, ApJ, 431, 137

Mutz, S. B., et al. 1994, ApJ, 434, L55.

Osterbrock, D. E., & Martel, A. 1993, ApJ, 414, 552

Postman, M., & Lauer, T. 1995, ApJ, in press

Press, W. H., Flannery, B. P., Teukolsky, S. A., & Vetterling, W. T. 1986, Numerical Recipes: The Art of Scientific Computing (Cambridge: Cambridge University Press), 289

Robinson, L. B., & Wampler, E. J. 1972, ApJ171, L83.

Romanishin, W., & Hintzen, P. 1989, ApJ, 341, 41

Schechter, P. 1976, ApJ, 203, 297

Smith, E. P., et al. 1986, ApJ, 306, 64

Stockton, A. 1978, ApJ, 223, 747

Stockton, A., & MacKenty, J. W. 1987, ApJ, 316, 584

Trauger, J. T., et al. 1994, ApJ, 435, L3

Véron-Cetty, M. P., & Véron, P. 1993, A Catalogue of Quasars and Active Nuclei (Sixth Edition), ESO Scientific Report, No. 13

Véron-Cetty, M. P., & Woltjer, L. 1990, A&A, 236, 69

Wyckoff, S., Wehinger, P. A., & Gahren, T. 1981, ApJ, 247, 750

Yee, H. K. C. 1987, AJ, 94, 1461

---

FIGURE CAPTIONS

Fig. 1.— The Unprocessed Images of Eight Luminous Quasars. This figure shows the long-exposure images of the eight luminous quasars observed with WFC2 through filter F606W. No luminous host galaxies are apparent.

Fig. 2.— The Two-dimensional F606W PSF. The figure shows the empirical point spread function generated from the combined images of three exposures of the star F141 in M67. The exposure times were 2.6 s, 12 s, and 70 s. The star was located 6″ from the center of chip 3 of WFPC2.

Fig. 3.— The Azimuthally-Averaged F606W PSF. The figure compares the azimuthallly-averaged Burrows (calculated) point spread function (solid line) and the azimuthal average of the empirical point spread function shown in the previous figure (dashed line).

Fig. 4.— The eight quasars with a best-fit stellar PSF subtracted. This figure shows the "long" (1400 s for PG 1307+085; 1100 s for the others) F606W observation of each of the eight quasars discussed in this paper. A best-fit stellar image from the M67 calibration data has been subtracted from each quasar image. Each of the panels is $20'' \times 20''$; the image scale is $0.0966''$ pixel $^{-1}$. Note that $1'' \approx 2$ kpc $(0.2/z)$.

Fig. 5.— HST images of eight galaxies. This figure shows the eight galaxies, taken from the fields of the quasar WFC2 observations, that were used to determine the sensitivity of the HST images to different types of quasar host galaxies. Each panel is 12.5″ on a side.

Fig. 6.— This figure shows the typical dependence of $\chi^2$ versus apparent magnitude of a putative host galaxy. In the particular case illustrated here, the S0/Sa galaxy shown in Figure 5e was added to the quasar PG 0953+414.

Fig. 7.— Close Companions of PKS 1302−102. This figure, a soft stretch of the 1100 s exposure of the PKS 1302−102 field, shows clearly two companion galaxies at separations of 1.1″ and 2.3″ from the quasar center. Their apparent magnitudes are, respectively, 20.6 and 22.0 mag. If the companions are at the distances indicated by the quasar redshift, then the projected separations from the quasar center-of-light are only 3 kpc and 6 kpc.

Fig. 8.— Comparison of HST and Ground-based (Palomar) Point Spread Functions. This figure compares the azimuthally-averaged radial surface brightness profile of the PSF of the F606W stellar HST images with the PSF obtained at the Hale Telescope with a V filter in 0.87″ (FWHM) seeing. Both the (Burrows) phenomenological and the measured stellar PSF are shown for the HST data.

Fig. 9.— Star PSF subtracted from different stellar images. Each of the panels shows the result of subtracting the combined stellar PSF shown in Figure 2 from the image of a different star. The first six panels, a-f, were obtained by subtracting the combined PSF of Figure 2 from stars observed with the WFC2 and F606W. The last three panels, g-i, show the results of subtracting the Figure 2 PSF from stellar images observed, respectively, with the F814W, F658N, and F439W filters.

Fig. 10.— Simulated AGNs. The figure shows a simulated AGN created from a bright stellar PSF (for the AGN nucleus) and a simulated nearly edge-on disk galaxy with an exponential scale height of 1.5″. The stellar PSF is taken from a slightly saturated star located 14″ from the center of CCD 2. The pseudo host galaxy has an apparent magnitude of 18.2 in the upper left panel and an apparent magnitude of 19.7 in the lower right panel. The limit of detectability with our techniques is 19.7 mag.